# Materials Screening Approach to
# Thermochemically Stable Thin Film Optical Emitters for Thermophotovoltaics

Declan Kopper and Marina S. Leite

Department of Materials Science and Engineering, University of California, Davis, CA

Corresponding author: mleite@ucdavis.edu

Abstract:

Thermophotovoltaics (TPVs) have the potential to exhibit higher power conversion efficiencies than traditional photovoltaics (PVs), with a broad range of applicability from waste recovery systems to aerospace solutions. They operate by preferentially radiating above bandgap photons *via* a high temperature optical emitter, whose spectrum is tuned through choice of materials and geometry. For TPV to be practically implemented, the emitters must be designed with a simple optical structure while remaining thermally stable. Here, we demonstrate coating/substrate bilayer thin films as a solution to these design criteria. With the optical data of 53 high melting point materials, we simulate the bilayer emissivity as a function of coating thickness for each thermochemically stable emitter operating at 1,800 °C. Emitter-cell systems are characterized by the cell power density and TPV conversion efficiency, constituting a TPV performance metric space. For a given bilayer and bandgap these coating thickness parameterized figures of merit form a performance metric curve, with the best points defining a tradeoff zone. We screen the resulting performance metric curves, identifying trends based on optical properties of the system, finding a high degree of tunability through the material selection step. For GaSb cells, >49% efficiency is achieved using AlN/W, a 5.6% increase over bulk W. By calculating the figures of merit for all systems with varying bandgap, we find unique emitter choices per PV cell for achieving the highest potential efficiency. Our materials screening approach uses physical data to identify improvements to emitters for experimental TPV designs.

**TOC:**

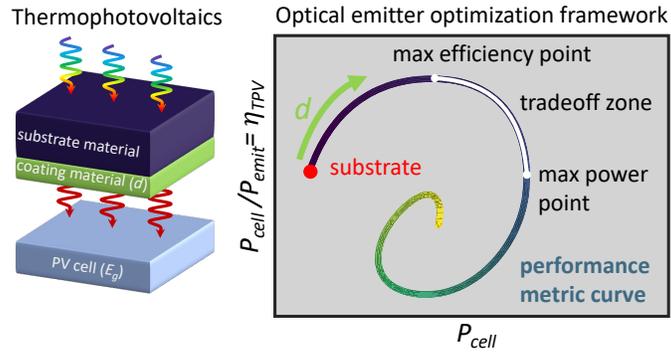

**Introduction**

      Photovoltaics (PVs) function through the photovoltaic effect, wherein photons with energy above the PV bandgap generate electron hole pairs which are converted to electrical energy. The efficiency of traditional PV, defined by the ratio of generated electrical power to incident solar power, is limited by losses from the broad blackbody spectrum of solar energy[1]. The largest loss comes from photons beneath the bandgap which are unable to contribute to the generation of electrical energy and are instead dissipated as heat[2]. Thermophotovoltaics (TPVs) offer a way to tune the spectrum incident on a PV device through a selective optical emitter[3]. Instead of energy coming from the sun, this component is heated such that it emits photons which are absorbed by the cell[4]. This process allows many types of sources to supply energy, making it useful for waste recovery and aerospace applications[5,6]. Selective optical emitters are designed such that these sub-bandgap photons are not radiated, suppressing this wasted energy[7]. Like solar PV, the TPV system can be characterized by two performance metrics: $P_{cell}$, the cell power density generated by the PV material, and $\eta_{TPV}$, defined as the ratio of cell power density to power density radiated by the emitter ($P_{cell}/P_{emit}$)[8].

      The design criteria of an optical emitter for TPVs require it to be stable at its operating temperature (typically above 1,000 °C)[9], with an emission spectrum that is tailored for usage by a given bandgap. Moreover, a simple structure makes it suitable for large-scale applications. Several methods for selective emitters have been proposed such as photonic crystals[10] and metamaterials[11], but they can suffer from stability issues under these high operating temperatures and scalability

challenges given their complex designs[12]. Thin-film optical emitters offer the opportunity to design multilayer structures with a tunable emission based on selection of thermochemically stable materials and their optical properties[13–15]. By choosing the number of layers, along with materials and thicknesses of each layer, the emissivity is uniquely defined[16]. These parameters can be tuned so that the radiated spectrum of the emitter operating at a given temperature is matched to the bandgap energy of the PV device[17]. A simple bilayer structure consisting of a thin coating layer and substrate can significantly improve performance compared to bulk emitters such as refractory metals and high temperature carbides which are typically used in proof-of-concept experiments[18]. Recently, remarkable progress has been achieved in the realm of experimental demonstrations using bulk emitters, with reports of achieving a $\eta_{TPV}$ of >38% with a graphite emitter operating at >1,800 °C[19], >40% with a 2,000 °C W emitter[20], and up to 43.8% with a 1,435 °C SiC emitter[21]. Yet, a limited choice of materials has been implemented as emitters. Therefore, there is a need to identify materials and their subsequent configurations that could lead to substantial gain in $P_{cell}$ and/or $\eta_{TPV}$. Nonetheless, there is a large parameter space to be considered even for a bilayer search, including the continuous variables of emitter temperature, the cell bandgap, and layer thicknesses, in addition to discrete variables of material selection (i.e., choice of *n,k*) for each layer.

Here we present a universal material screening paradigm to identify and quantify the best material choices for TPV coating/substrate optical emitters. By using >50 high-melting-temperature materials, we calculate the emissivity of all bilayers operating at 1,800 °C as a function of coating thickness *d*. We focus on material pairs that can form thermally stable interfaces. For a given bilayer emitter and PV, the TPV is assessed by our figures of merit $P_{cell}$ and $\eta_{TPV}$, defined as existing in a performance metric space. The combination of intrinsic optical properties for each layer and bandgap produces a performance metric curve parameterized by the thickness of the coating (*d*). The resulting performance metric curve encompasses tradeoff zones between the maximum efficiency point (MEP) and maximum power point (MPP). A selection of calculated performance metric curves is shown to emphasize the variability in shape that occurs based on material type. Using a bandgap dependent search, we down select notable emitters and display their performance operating with a GaSb PV cell, showing that refractory metals with dielectric coatings such as AlN/W demonstrate > 5.0% improvement over a bulk W. The performance metric curve is then calculated for each emitter over a bandgap range of 0.5 eV to 1.2 eV. Using the previously chosen emitters we emphasize the bandgap dependence of the curves, and that bilayers have a specific bandgap under which they achieve the highest MEP. We conclude by highlighting the MEP per bandgap value of these selected emitters, showing the gain in efficiency achieved by these emitters compared to a pure blackbody source, and that the best performing emitter varies per bandgap. Our approach surveys a large array of materials combinations, merging physical data with simulations to identify alternative, scalable optical emitters to implement in TPVs.

**Results and Discussion**

*Optical Emitters Design and the Performance Metric Space*

The basis of our materials' screening approach lies in the selection of the emitter coating and substrate materials using a scalable configuration, shown in Figure 1(a). The refractive index ($n,k$) of the chosen materials and coating thickness ($d$) results in a tuned emission[22]. When a PV cell with a given bandgap $E_g$ absorbs the emission as seen in Fig 1(a), the emitter-cell system is then defined by two figures of merit: the cell power density ($P_{cell}$) and the TPV efficiency ($\eta_{TPV} = P_{cell}/P_{emit}$) (for additional information on the calculation of emissivity and TPV performance, see the supplemental file). By selecting the emitter materials and the PV bandgap, we calculate a performance metric curve in this space which is parameterized by the coating thickness $d$.

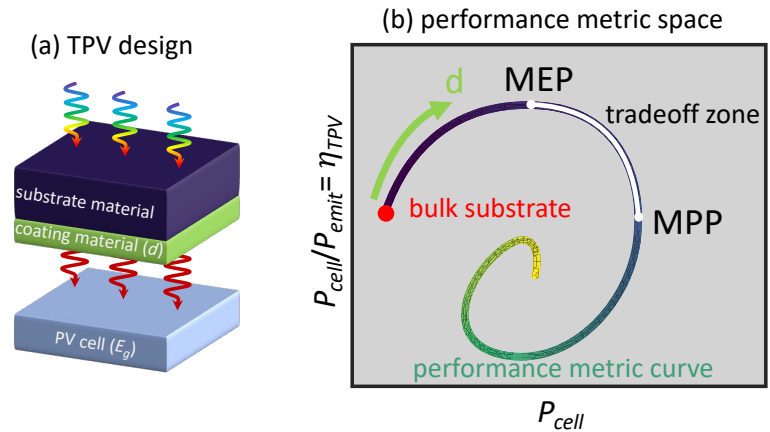

**FIG. 1.** Performance metric space concept for bilayer emitters. (a) Schematic of thermophotovoltaic (TPV) design consisting of a bilayer optical emitter with coating layer of thickness $d$ (nm) and a bulk substrate. Energy is transferred to the emitter, which selectively emits to the photovoltaic cell of bandgap $E_g$ (eV). (b) Performance metric space for thermophotovoltaic figures of merit, cell power density ($P_{cell}$) and TPV efficiency ($\eta_{TPV}$). In color: performance metric curve based on the TPV design in (a), where the figures of merit are parameterized by coating thickness. Tradeoff zone (white) consists of points with power greater than the maximum efficiency point (MEP) $P_{cell}$ and efficiency greater than the maximum power point (MPP) $\eta_{TPV}$.

Fig 1(b) displays a schematic performance metric curve for a general coating/substrate pair and PV cell. As a reference, we display the performance for the bulk substrate material without any coating in red. Adding a coating material and linearly varying its thickness (from 1 nm to 1000 nm) is equivalent to moving along the performance metric curve, which shows a highly non-linear behavior. Two points along the curve are identified in white, one belonging to the maximum efficiency point (MEP), where $\eta_{TPV}$ is maximized on the curve, and the other to the maximum power point (MPP), where $P_{cell}$ is maximized on the curve. Any points with power above the MEP power and efficiency above the MPP efficiency are part of a 'tradeoff zone', shown in white. This section is highlighted to denote the tradeoff that exists when designing a TPV system. Typically, the MEP corresponds to a highly suppressed emission window for below $E_g$ photons which reduces $P_{emit}$ but does not improve $P_{cell}$. Conversely, the MPP corresponds to a high broadband emissivity that improves the number of photons converted into $P_{cell}$ while consequently increases $P_{emit}$. Ideally, the two points (MEP and MPP) would coincide, but an ideal step function for emissivity is difficult to achieve, especially with few layers[23]. Note that even if there is only a marginal improvement of either figure of merit at the MEP or MPP compared to other points on the curve, or a point inside the zone that is outperformed in either metric by another point, it is still considered as part of the tradeoff zone. While others have defined a new figure of merit based on these metrics[24,25], we do not define a tradeoff function between the two figures of merit. Selections of best performing emitters will be done based on $\eta_{TPV}$, as TPVs have the advantage of achieving higher efficiencies than traditional PV which is limited by the fixed incident power from the solar spectrum[26]. Yet, the independent assessment of the conditions required for MPP is relevant in situations where high absolute power is preferred or needed. Note that our approach is material agnostic and can be implemented to any TPV design, making it universal.

The following sections of this Article discuss: (i) how and why the performance metric curve can be engineered, (ii) select high performing thermally stable material combinations operating with a GaSb cell, (iii) the effects of the PV cell bandgap ($E_g$) on the overall performance metric space, (iv) the ideal optical emitters for common TPV cells: InGaAsSb, InGaAs, Ge, GaSb, and Si. Throughout, we discuss the potential gain when optimizing $P_{cell}$ and $\eta_{TPV}$ independently, which can be implemented as a guideline in the search for optical emitters that can surpass the current state-of-the-art.

*Material Screening for Optical Emitters*

Using the refractive index data of 53 materials with melting points >2,000 °C, we calculate the emissivity of all possible bilayers over varying coating thickness. We choose to exclusively consider the emitters whose coating and substrate materials are expected to be thermochemically stable at 1,800 °C[3], leaving 200 potential bilayers. For comparison, we also calculate the emissivity of each material as a bulk emitter, depicting how much change is achievable by the introduction of the coating layer with contrast in refractive index. With the 1,800 °C emitter spectrum incident on the PV material (and thus, bandgap) of choice, we calculate the thickness dependent performance metric curve for that system. To show the variability that occurs in performance metric curves based on materials selection, Figure 2 displays an assortment of curves with different stable emitters, all for a GaSb ($E_g$ = 0.73 eV) PV cell, commonly used for thermophotovoltaics[27]. The set of emitters shown here is chosen to emphasize trends that can be seen throughout all performance metric curves (for additional data see Figure S1-S6 in the supplemental file). Note that unlike the schematic shown in Fig 1(b), which has a continuous tradeoff zone, some emitters here have multiple tradeoff zones, meaning there is an intermediate range of coating thickness values which lead to lower figures of merit between the MEP and MPP. Additionally, some plots do not contain the substrate performance metric point (in red), which will be discussed later. The variety of shapes demonstrates the high degree of TPV performance tunability by changing the materials comprising the emitter, and the differing effect a change of coating thickness $d$ can have on the overall TPV performance.

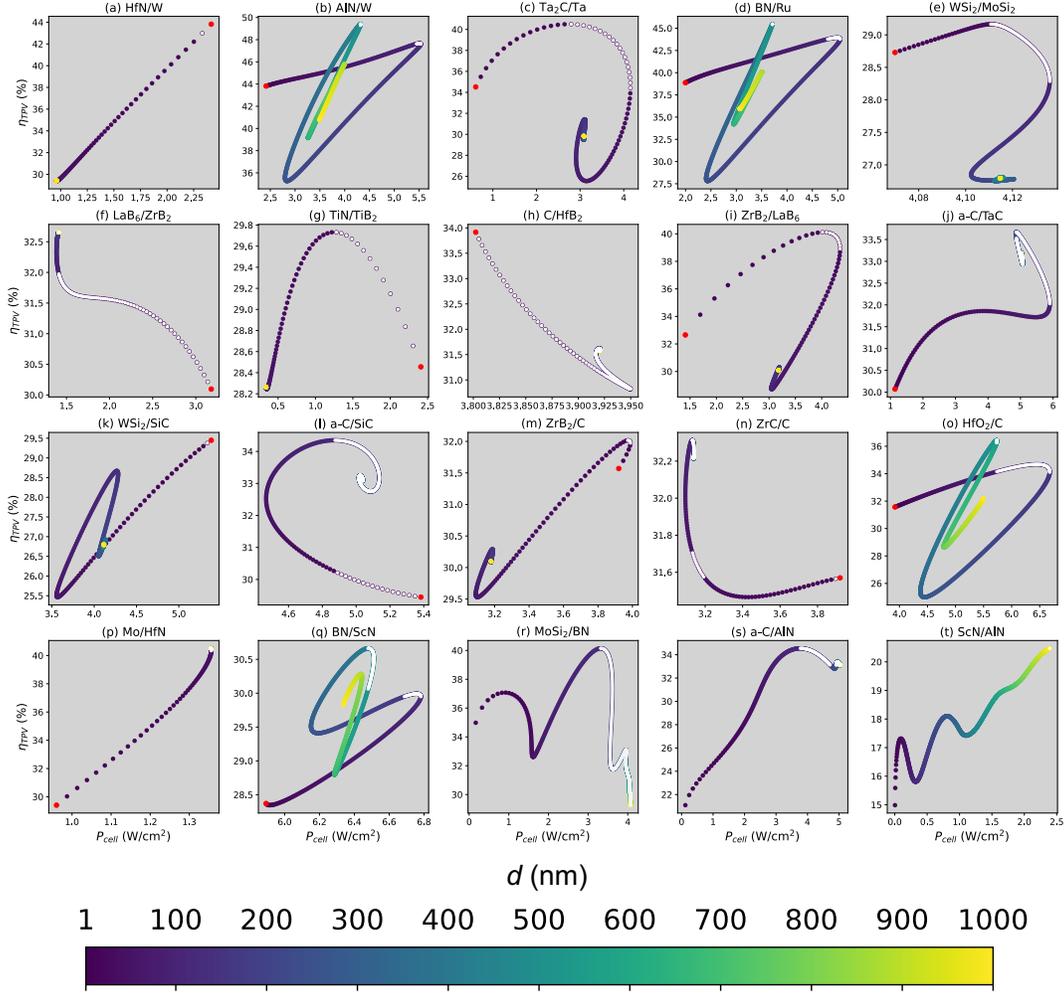

**FIG. 2.** Performance metric curves for selected emitters using a variety of coating/substrate pairs with a GaSb ($E_g$ = 0.73 eV) photovoltaic cell. The color gradient of each curve refers to the coating thickness variation $d$, from 1 nm to 1000 nm. White points and lines correspond to the tradeoff zone and the red point indicates the performance of the bulk substrate ($d$ = 0 nm).

Here, we highlight select performance metric curves, emphasizing the origins of the stark differences between them based on the behavior of $n,k$. A trend that separates the curves is based on the type of each material in the bilayer (and, thus, their optical properties). Broadly, the 53 materials can be split into three optical classes, those considered to be absorbing ($k \neq 0$), semi-absorbing ($k \approx 0$), and transparent ($k = 0$) over the wavelength range investigated. The bilayer behavior is dependent on the combination of optical layers. The first emitter class, seen in Figure 2(a), (c), (e-n), (p), (r), and (s), are those whose coating layer is an absorbing material, shown by curves which are primarily purple (i.e., ultra-thin coating thickness) and converge to a singular yellow point (with $d$ = 1000 nm thick coating). In this situation, only a thin layer of the coating is required for the coating layer to become opaque. Within this thickness regime prior to converging, there is varied behavior depending on substrate and coating materials (refractory metal, nitride, carbide, boride, etc.), with one or two tradeoff zones of varying thicknesses ranges. The second emitter class, shown in Figure 2(b), (d), (o), and (q) all consist of a transparent dielectric coating layer on a substrate made of an absorbing or semi-absorbing material. The general shape of the performance metric curve is denoted by a spiraling shape, with two separate tradeoff zones containing either the MEP or MPP. This curling occurs because the transparent coating does not absorb light and thus will not converge to bulk, allowing for tunability of the substrate by optically impedance matching different wavelengths depending on coating thickness. For Figure 2(t), consisting of a semi-absorbing ScN coating on a transparent AlN substrate, there is a weaker thickness dependence approaching bulk ScN compared to the other emitters. Because the coating thickness has a maximum of 1000 nm, increasing this value would lead to a further segment of performance metric curve until a sufficient thickness where the coating is opaque. Generally absorbing and semi-absorbing should be treated as part of a spectrum, but here the maximum coating thickness of 1000 nm artificially distinguishes the two classifications. As noted earlier, Figure 2(r), (s), and (t) all lack the point for the bulk emitter as the substrates are transparent materials and would not operate as bulk. Another class of emitters consisting of two transparent materials are not shown, as this combination would also lead to emitters with very modest performance. This distinction of the four bilayer optical classes helps us in identifying how the materials selected can present similar behaviors while still containing a high degree of uniqueness and, in some cases, tunability.

To compare the dependence of emitter materials on TPV performance, we choose specific coating/substrate pairs in Fig 2 to highlight these aspects. For example, Fig 2(f) and (i) depict $LaB_6$ and $ZrB_2$ as coatings, respectively, with the opposite material as the substrate. Note that while the endpoints of both curves are of bulk $ZrB_2$ and bulk $LaB_6$ (see Figure S7 for comparison of the bilayers in Fig. 2 on equivalent axes), the $ZrB_2$/$LaB_6$ emitter in (i) performs better, with a tradeoff zone with superior values to any points in the tradeoff zone of the emitter in (f). Another comparison can be drawn between Fig 2(k) and (l), which both consist of a SiC substrate with different coating layers. In (k), $WSi_2$ leads lower metrics throughout the curve compared to bulk SiC, whereas in (l) the addition of a-C results in a tradeoff zone containing the MEP. Finally, Figure 2(j) and (s) both have a-C as a coating layer, on a TaC substrate and a AlN substrate, respectively.

Overall, note that most emitters outperform their pure substrate counterparts with the addition of a coating layer with optimized thickness to maximize either power, efficiency, or both.

*Material Down Selection with GaSb Cells*

With the performance metric curve of a bilayer defined at a specific bandgap, we calculate the results for each emitter over $E_g$ = 0.5 – 1.2 eV in 0.05 eV increments. By introducing $E_g$ as another variable, we identify material systems that work with other PV cells, beyond GaSb. With a low bandgap cell, less of the emitter spectrum is sub-bandgap. Consequently, there is higher in-band usage, but simultaneously less effective use of high energy photons. Bandgap materials such as Si are more ubiquitous but lead to less in-band emission and thus reduced power, requiring reflection of the out-of-band photons back to the emitter[28]. Here, we focus on bilayers that have high MEPs relative to other emitters with the same $E_g$, as efficiency is the more widely considered metric for TPVs. The MEP efficiency point as a function of bandgap for all 200 stable bilayer emitters and 53 bulk emitters can be found in Figure S8, which was used for assessment of the emitters.

Figure 3(a-d) shows the performance metric curves for a selection of emitters with the highest MEP for GaSb cells: AlN/W, MgO/W, SiC/LaB$_6$, and AlN/Mo, respectively. The emitters B$_4$C/HfB$_2$ and Cr$_2$O$_3$/ZrO$_2$ in Fig 3(e) and (f) are picked because they perform well with our lowest and highest investigated bandgaps, respectively. Oxides are used with high bandgap materials because their transparency over a large wavelength range suppresses unusable energy for the cell[9]. Overall, the presented results reflect experimental bulk emitter choices while demonstrating the gain in performance achieved by moving from a bulk to bilayer emitter with tunable coating layer. In experiments involving TPV, refractory metals such as W and Mo are often used as bulk emitters because of their natural spectral emissivity[29], with experiments also using anti-reflective coating layers to tune the emissive properties of the structure[30]. SiC, also used as a bulk emitter in experiments[20], is shown here to have improved performance when coated as a thin film layer.

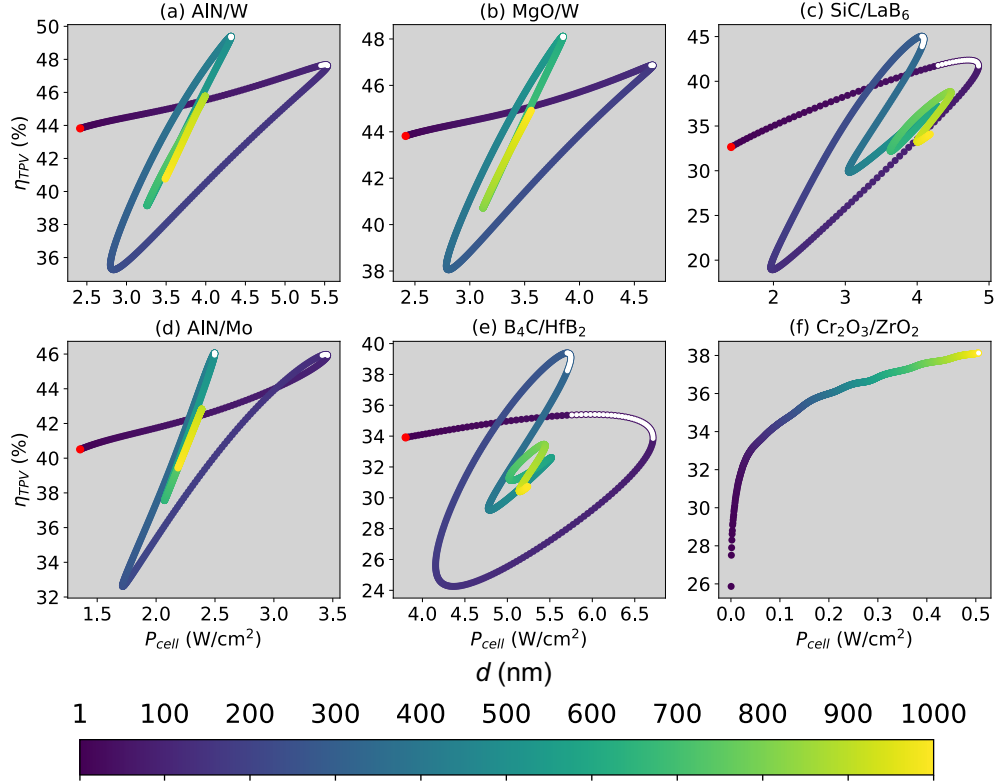

**FIG 3.** Performance metric curves of efficiency $\eta_{TPV}$ as a function of $P_{cell}$ for emitters. (a) AlN/W (b) MgO/W (c) SiC/LaB$_6$ (d) AlN/Mo (e) CrO$_2$O$_3$/ZrO$_2$ (f) B$_4$C/HfB$_2$ with a GaSb PV ($E_g$ = 0.73 eV). The color scale refers to thickness variation of the coating layer. White points and lines correspond to the tradeoff zone and the red point indicates the performance of the bulk substrate ($d$ = 0).

As done previously, it is useful to draw comparisons between the emitter selections, noting the important properties that lead to their high performance. AlN/W in Fig 3(a) has the highest MEP for any GaSb TPV system with an efficiency of 49.4%. The bilayer is composed of a W substrate, which in bulk has an efficiency of 43.8%, and a transparent coating AlN whose thickness can be optimized to reduce any sub-bandgap emission. Similarly, MgO/W in Fig 3(b) has the third highest MEP for GaSb, replaces AlN with MgO, achieving a similar degree of tunability due to the transparent oxide. CaO/W, see Fig S8, has the second highest MEP, enforcing this trend that dielectric coatings to W lead to high emissivity control and thus high efficiency. Now, comparing AlN/W with Fig 3(d), AlN/Mo has the fifth highest GaSb MEP, here replacing the substrate with a different refractory metal. The strength of the broad screening approach is that a wide array of material options for coatings can be tested, demonstrating here that AlN is well matched for both W and Mo, and these two metals outperform other emitter candidates.

The high performance of refractory metals with dielectric coatings is due to the spectral emissivity trend, being high for short wavelengths and low for long wavelengths relative to the material bandgap (see Fig S9 for emissivity profiles). The addition of the coating layer tunes this profile, with the coating MEP thickness further suppressing out-of-band emission, and the MPP coating increasing broadband emission. The distinct tradeoff zones indicate a thickness that leads to maximum in-band emissivity and out-of-band suppression does not exist. Compare this with the profiles of SiC/LaB$_6$ and B$_4$C/HfB$_2$, which achieve higher broadband emissivity across the entire wavelength spectrum. As seen in Fig 3(c) and (e), these bilayers do not reach the same efficiencies as the (a), (b), or (d), but do achieve relatively high MPPs. In contrast, Cr$_2$O$_3$ only emits at short wavelengths, leading to high efficiencies but minimal power. As the coating thickness is increased, the in-band emissivity rises, enhancing both power and efficiency, as seen in Fig 3(f). The tunability arises from the material selection ($n,k$) and coating thickness optimization, but here we have defined in- and out-of-band emissivity relative to GaSb, which highlights the importance of bandgap selection on TPV performance.

*Bandgap Effects on TPV Performance Metric Curve*

To resolve the effects of the PV $E_g$ on the overall performance metric curve, Figure 4 shows the curves for the previous six emitters, with the color of each curve representing the bandgap in 0.1 eV increments and the corresponding darkness referring to the coating thickness variation ($d = 1 – 1000$ nm). Because $P_{emit}$ is independent of the bandgap, it is $P_{cell}$ that leads to the change in both figures of merit, determining the absorption cutoff of usable photons and defining in-band and sub-bandgap. The overall non-linear dependence of $P_{cell}$ and $\eta_{TPV}$ is evident, with differing behavior based on the optical class of the emitter. For AlN/W, MgO/W, and AlN/Mo (Fig 4(a), (b), and (d)), the trend in behavior between the three is overall similar because of the nature of the dielectric/metal structure (and their intrinsic $n,k$). Each bandgap curve has the characteristic spiraling shape as seen in Fig 3. However, for these three systems the low bandgap 0.5 eV curve only contains one tradeoff zone encompassing both the MEP and MPP. At 0.7 eV the shape is altered, with the larger coating thickness region of the curve outperforming the low thickness segment. Moreover, the MEP and MPP for the refractory metal coated emitters reach their global maximum at the same bandgap of 0.7 eV, with both figures of merit reduced with subsequent increases in the bandgap. This result shows that the spectrum emitted by bilayer refractory metals, is uniquely suited for PV materials with bandgaps near 0.7 eV (including GaSb), achieving an absorption cutoff that yields effective usage of high energy photons.

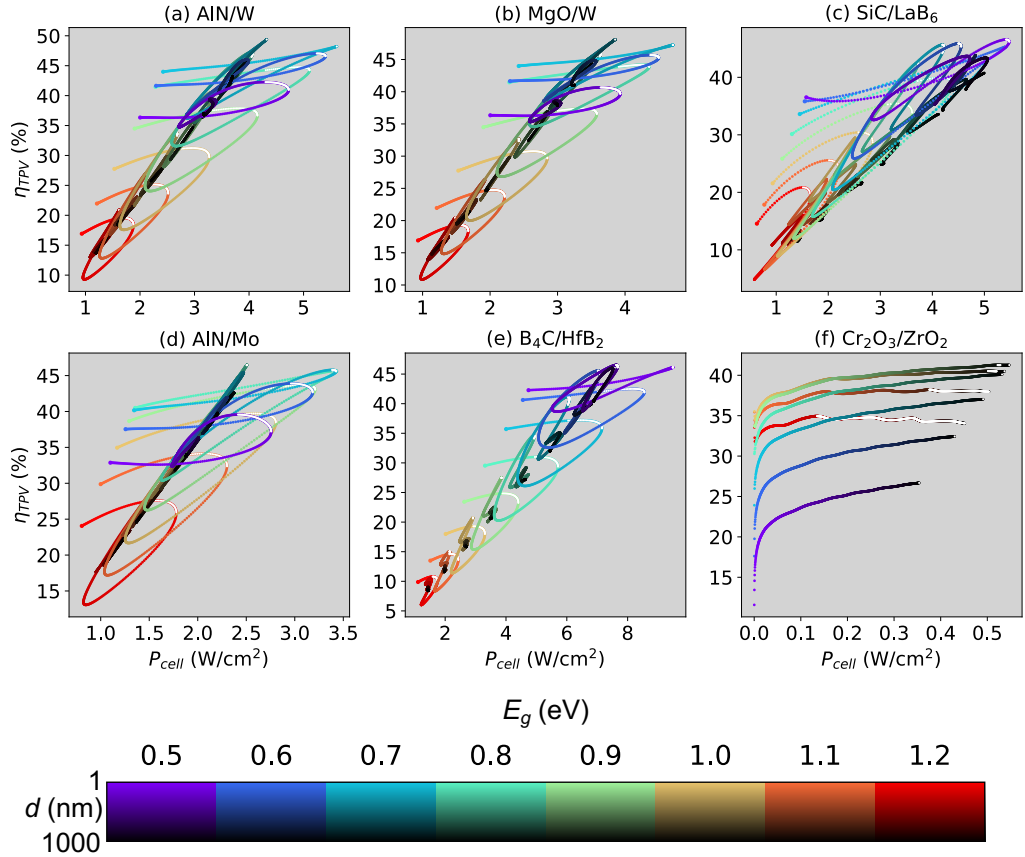

**FIG. 4.** Performance metric curves as a function of PV bandgap for (a) AlN/W (b) MgO/W (c) SiC/LaB$_6$ (d) AlN/Mo (e) B$_4$C/HfB$_2$ (f) Cr$_2$O$_3$/ZrO$_2$. The color scale refers to an $E_g$ variation from 0.5 eV to 1.2 eV. The darkness of the curve is related to the thickness of the coating layer, varying from 1 nm to 1000 nm.

Comparing the performance metric of emitters with a refractory metal substrate in Fig 4(a), (b), and (d) with SiC/LaB$_6$ and B$_4$C/HfB$_2$ in Fig 4(c) and (e), their global MEP occurs for 0.5 eV, with performance continually increasing with decreasing bandgap. This trend is due to their broadband emissivity, which at lower bandgaps yields a more in-band absorption which also increases the power, unlike the coating/refractory metals which have a turning point in functionality. In contrast, Cr$_2$O$_3$/ZrO$_2$ in Fig 4(f) shows a reversal of bandgap dependence, increasing both metrics until a global maximum at 0.9 eV and then a decrease. Because of the naturally suppressed emission of long wavelengths, reducing the in-band region to more effectively use the high energy photons increases the efficiency up until a diminishing point in the emission spectrum of an 1,800 °C emitter. Note again that regardless of bandgap, these emitters see an improvement in power and efficiency by using the bilayer structure with optimized thicknesses as opposed to a bulk substrate.

To summarize our findings of the TPV system's bandgap dependence seen in Fig 4, Figure 5 shows the MEP efficiency of the previous six emitters at each 0.05 eV incremental bandgap point, with a curve of the bandgap dependent efficiency expected for a blackbody emitter operating at 1,800 °C. Bandgaps of most the commonly used PV materials for TPV are highlighted to stress the importance of $E_g$ when selecting optical materials. The $\eta_{TPV}$ gain is evident when comparing the blackbody and the optimized optical emitter spectra. The most modest gain occurs for InGaAsSb (lattice-matched to GaAs), as the high broadband emissivity achieved by SiC/LaB$_6$ and B$_4$C/HfB$_2$ is like that of a blackbody but suppresses out-of-band emission. For wider bandgaps AlN/W is the best emitter option, achieving the highest MEP for InGaAs and Ge, besides GaSb, with increasing gain over the blackbody efficiency. It is also evident here (and through Fig S8) that an AlN/W emitter with a bandgap near GaSb achieves the highest MEP of all emitter-cell systems. Beyond 0.8 eV, the AlN/Mo bilayer outperforms AlN/W, working better for wider bandgap materials. Cr$_2$O$_3$/ZrO$_2$ reaches the highest MEP for Si (1.1 eV), also showing the largest improvement over the use of a pure blackbody source. These results quantify the benefit of searching for material systems beyond those implemented as emitters in TPV, with gain by using a simple bilayer structure. Experimental proof-of-concept realizations of TPV using refractory metals (W) and SiC represent an important step towards validating this renewable energy technology. Yet, by expanding the search for materials adopted as optical emitters in TPV there is an opportunity to significantly improve device performance without need for complex fabrication procedures or nanostructures.

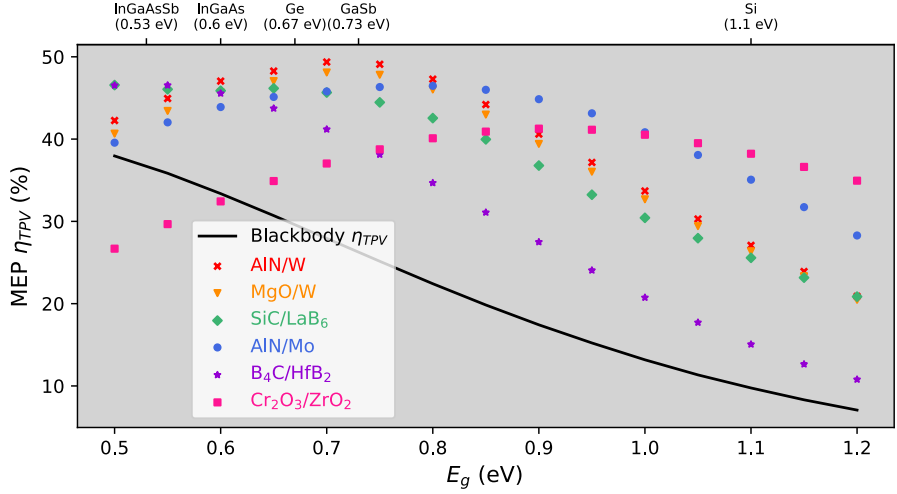

**FIG. 5.** The MEP $\eta_{TPV}$ for each of the previous six emitters as a function of photovoltaic cell bandgap. The black line shows the efficiency of an ideal blackbody as a function of bandgap. The mostly common semiconductors materials used for the photovoltaic cell are highlighted along the top axis.

**Conclusions**

We showed a universal, computational methodology to optimize the performance of thermally stable optical emitters for TPV, based on the screening 50+ materials and their thermochemically stable bilayer configurations. Given the selected emitter materials we identified how variations in the coating tailor the relationship between the two most important metrics that define performance in TPV: $\eta_{TPV}$ and $P_{cell}$. We used 20 bilayers operating with a GaSb cell to explain the overall behavior of the performance metric curve, highlighting the variation and tunability attained by using distinct classes of materials ($n,k$) for both the coating and the substrate. Top performing emitters were selected; with the best emitters for GaSb presenting two tradeoff zones in the performance metric curve based allowing for optimized power or efficiency. We showed the bandgap influence on the overall "trajectory" of the curves based on the emitter type. Another key finding was that the maximum $\eta_{TPV}$ per bandgap for the emitters, with AlN/W achieving a high performance over the investigated 0.5 – 1.2 eV range and reaching a global maximum efficiency of 49.4% with a GaSb cell. Overall, the five most adopted cells (InGaAsSb, InGaAs, Ge, GaSb, and Si) in TPV could significantly benefit from our optimization paradigm, with a maximum gain to ~50%. Our results are relevant for presenting simple improvements to emitter structures (beyond bulk SiC, W, etc.) with performance considerably above the current state of the art. Because we exclusively considered materials that were determined to be thermochemically stable up to 1,800 °C, the simulations provide a realistic pathway for the next generation of optical emitters.

**Acknowledgments**

The authors thank J.N. Munday for fruitful discussions. MSL thanks the financial support from UC Davis College of Engineering Next Level Research Vision. DK acknowledges funding from the Marjorie and Charles Elliot, Fletcher Jones, and George and Dorothy Zolk Fellowships.

**Author Declarations**

The authors declare no conflict of interest.

**Supplementary File**

*Emissivity Calculation*

Optical calculations were performed using the index of refraction ñ($\lambda$)=n($\lambda$)+ik($\lambda$) for 53 materials over a wavelength range of 350-3000nm using the data from Dias et al. The transfer-matrix-method (TMM) was used to determine the absorption of the thin film stack over this range. This method calculates the reflectance and transmittance of a multilayer thin film stack through input of the thickness and optical data of each layer. The python TMM package created by Steven Byrnes was used for all TMM calculations.

TMM was performed differently based on the properties of the underlying substrate layer, characterized as either absorbing or non-absorbing. Non-absorbing was defined here if the optical thickness for any wavelength of the material was less than 12:

$$\frac{4\pi k(\lambda)}{\lambda} d < 12$$

Where k($\lambda$) is the complex index of the substrate and d is the substrate thickness, taken as 0.1mm. The substrate was treated as semi-infinite for absorbing substrates to reduce computational time, with absorbance taken as A($\lambda$)=1-R($\lambda$). For non-absorbing substrates the 0.5mm substrate thickness was used and absorbance calculated as A($\lambda$)=1-R($\lambda$)-T($\lambda$), where T is transmittance. Coating layers were simulated with thicknesses of 1-1000nm. By Kirchhoff's law, the emissivity of the material is equal to the absorbance, $\varepsilon(\lambda)$=A($\lambda$). The optical data of each material is from a room temperature measurement, but taken to be equivalent to that which would occur in the device operating at 1,800 °C. Additionally, the coverage of the 1,800 °C blackbody energy spectrum for our wavelength range is 75.5%. These two factors stress a need for measurements of material optical properties at both elevated temperatures and in the IR wavelength range.

*Thermal Stability Determination*

When operating at elevated temperatures, the thermally stability of the emitter must be considered. The individual materials were selected to have melting temperatures higher than 2000°C, but multilayer emitters must also meet the criteria that the material system is stable. This means the materials must not form a new phase at the interface, which would lead to different properties in the emission spectrum. Stability determinations were taken from Dias et al, which used the phase diagrams of each system to identify the state at 1,800 °C. Only those listed as stable were considered for the down selection steps.

*TPV Calculation*

The emissivity is used for the calculation of the JV characteristic of the solar cell using the diode equation:

$$J(V) = J_L - J_0(e^{\frac{qV}{kT_c}} - 1)$$

Here, $J$ is the cell current density, $q$ is the electron charge, $V$ is the cell voltage, $k$ is the Boltzmann constant, and $T_c$ is the cell temperature taken to be 300K. $J_L$ is the light generated current density given by

$$J_L = q \int_{350}^{3000} \varepsilon(\lambda) \bar{n}_B(\lambda, T_E) IQE(\lambda) d\lambda$$

And $J_0$ is the dark saturation current density

$$J_0 = q \int_{350}^{3000} \bar{n}_B(\lambda, T_c) IQE(\lambda) d\lambda$$

where $\bar{n}_B(\lambda, T)$ is the blackbody photon spectral radiance, $T_E$ is the emitter temperature, and IQE is the internal quantum efficiency. The IQE is taken to be an ideal step function at the bandgap of the cell, with all photons of higher energy than the bandgap being absorbed. $P_{cell}$ is then calculated from the maximum power point of the diode equation. The efficiency is defined as

$$\eta_{TPV} = \frac{P_{cell}}{P_{emit}}$$

With $P_{emit}$, the energy density emitted by the emitter

$$P_{emit} = \int_{350}^{3000} \varepsilon(\lambda) B(\lambda, T_E) d\lambda$$

Where $B(\lambda, T)$ is the blackbody energy density.

*Supplemental Figures*

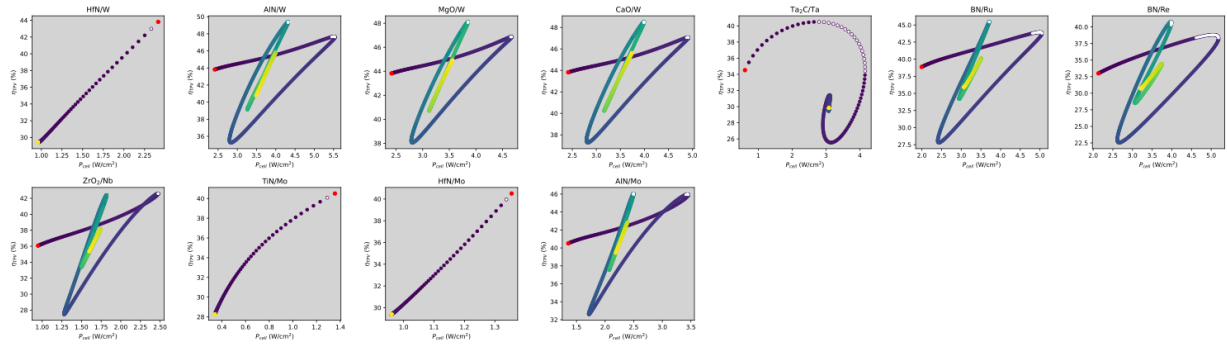

**Figure S1** Performance metric curves for TPV systems with a GaSb PV cell and bilayer emitters with a metal substrate. The background color of each graph refers to a distinct substrate material. White line and red dot correspond to the tradeoff zone and the performance of the bulk substrate, respectively.

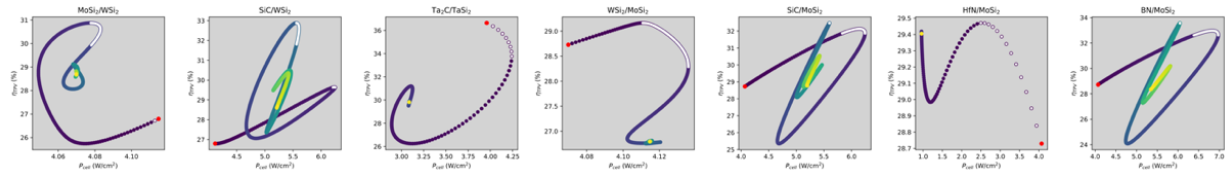

**Figure S2** Performance metric curves for TPV systems with a GaSb PV cell and bilayer emitters with a silicide substrate. The background color of each graph refers to a distinct substrate material. White line and red dot correspond to the tradeoff zone and the performance of the bulk substrate, respectively.

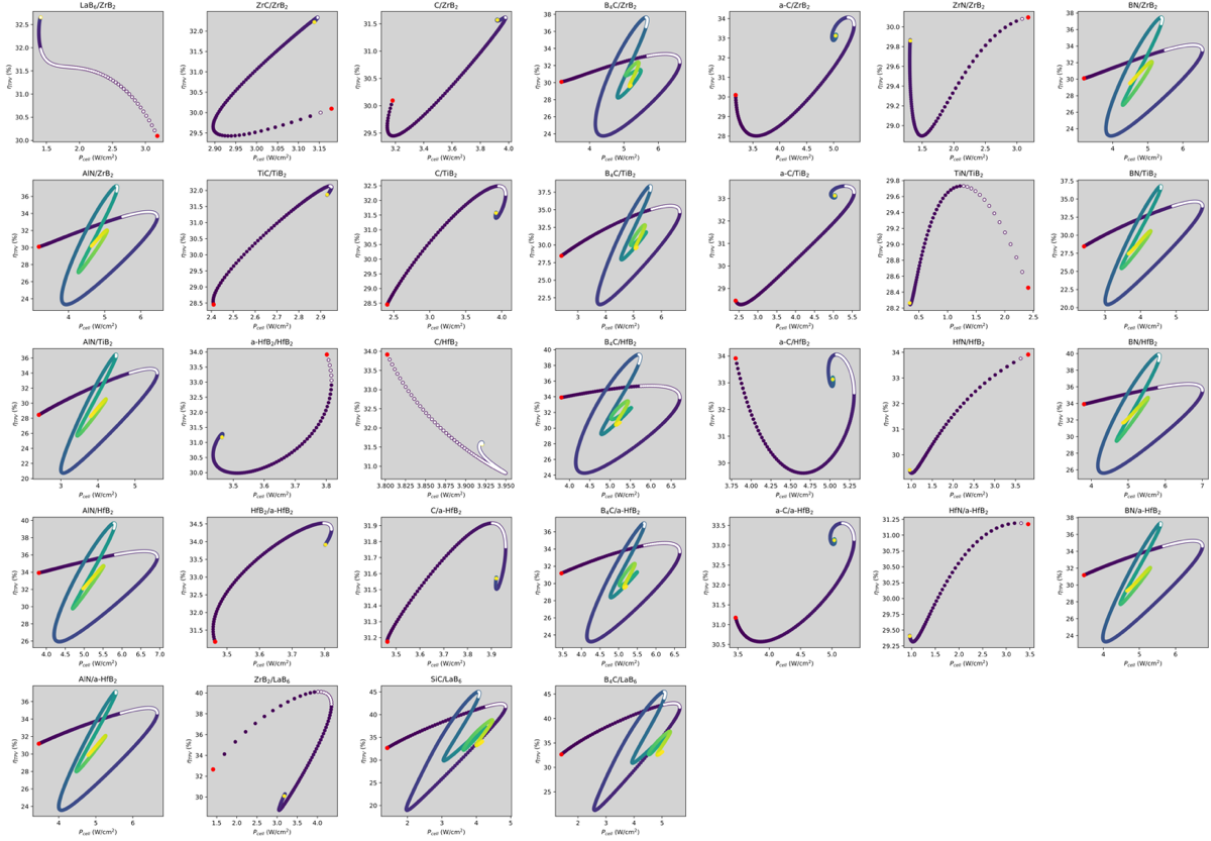

**Figure S3** Performance metric curves for TPV systems with a GaSb PV cell and bilayer emitters with a boride substrate. The background color of each graph refers to a distinct substrate material. White line and red dot correspond to the tradeoff zone and the performance of the bulk substrate, respectively.

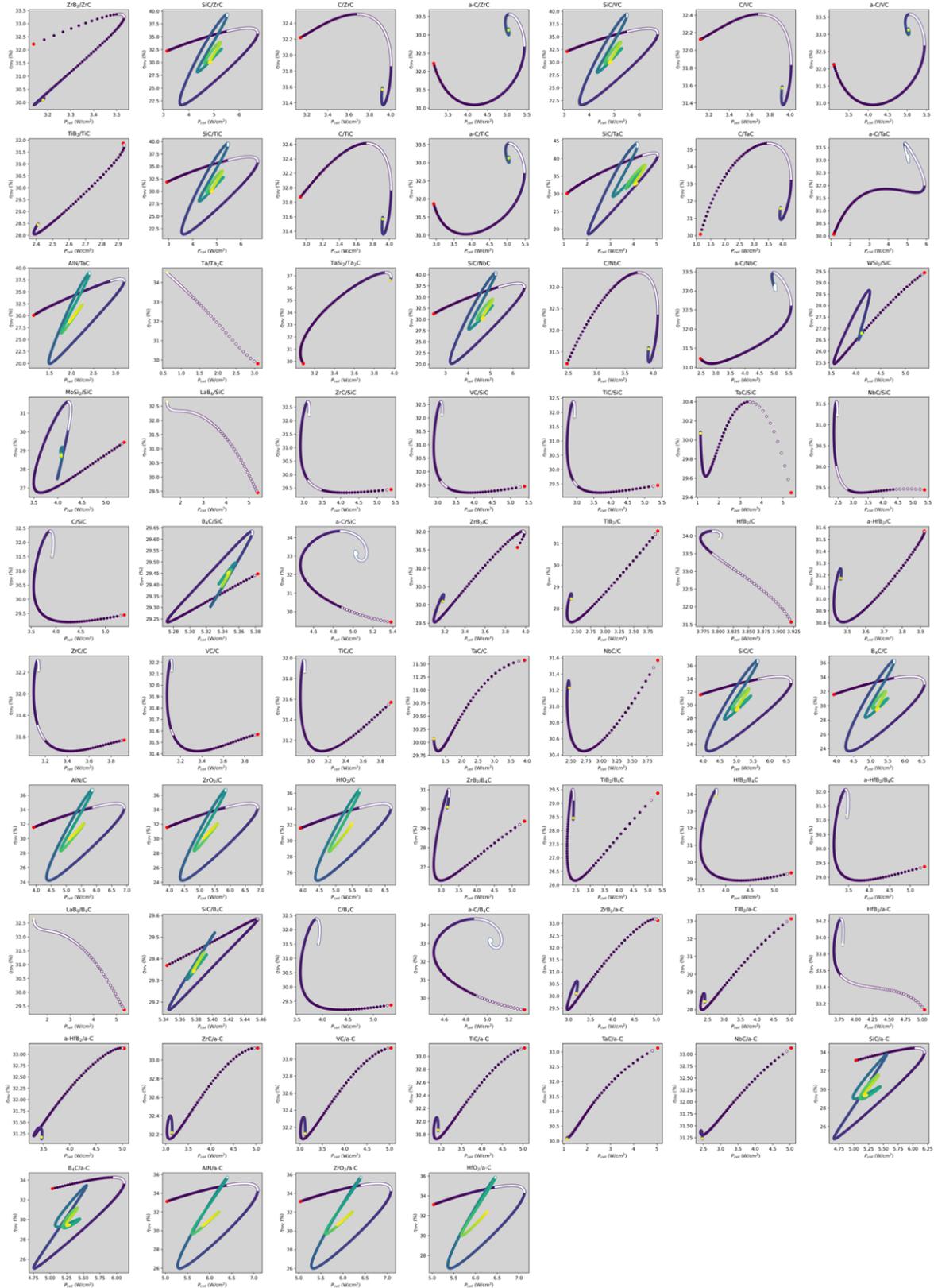

**Figure S4** Performance metric curves for TPV systems with a GaSb PV cell and bilayer emitters with a carbide substrate. The background color of each graph refers to a distinct substrate material.

White line and red dot correspond to the tradeoff zone and the performance of the bulk substrate, respectively.

**Figure S5** Performance metric curves for TPV systems with a GaSb PV cell and bilayer emitters with a nitride substrate. The background color of each graph refers to a distinct substrate material. White line and red dot correspond to the tradeoff zone and the performance of the bulk substrate, respectively.

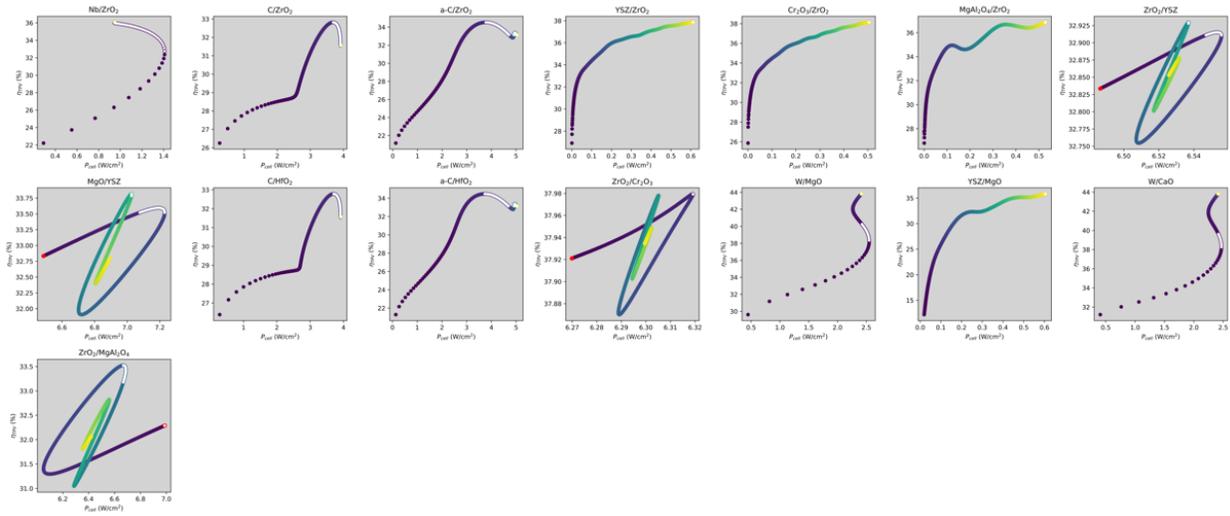

**Figure S6** Performance metric curves for TPV systems with a GaSb PV cell and bilayer emitters with an oxide substrate. The background color of each graph refers to a distinct substrate material. White line and red dot correspond to the tradeoff zone and the performance of the bulk substrate, respectively. Several curves for stable bilayers yielded MPPs of $P_{cell}$ < 0.2 W/cm$^2$ and their curves are not shown, but their MEP data can be found in Figure S9.

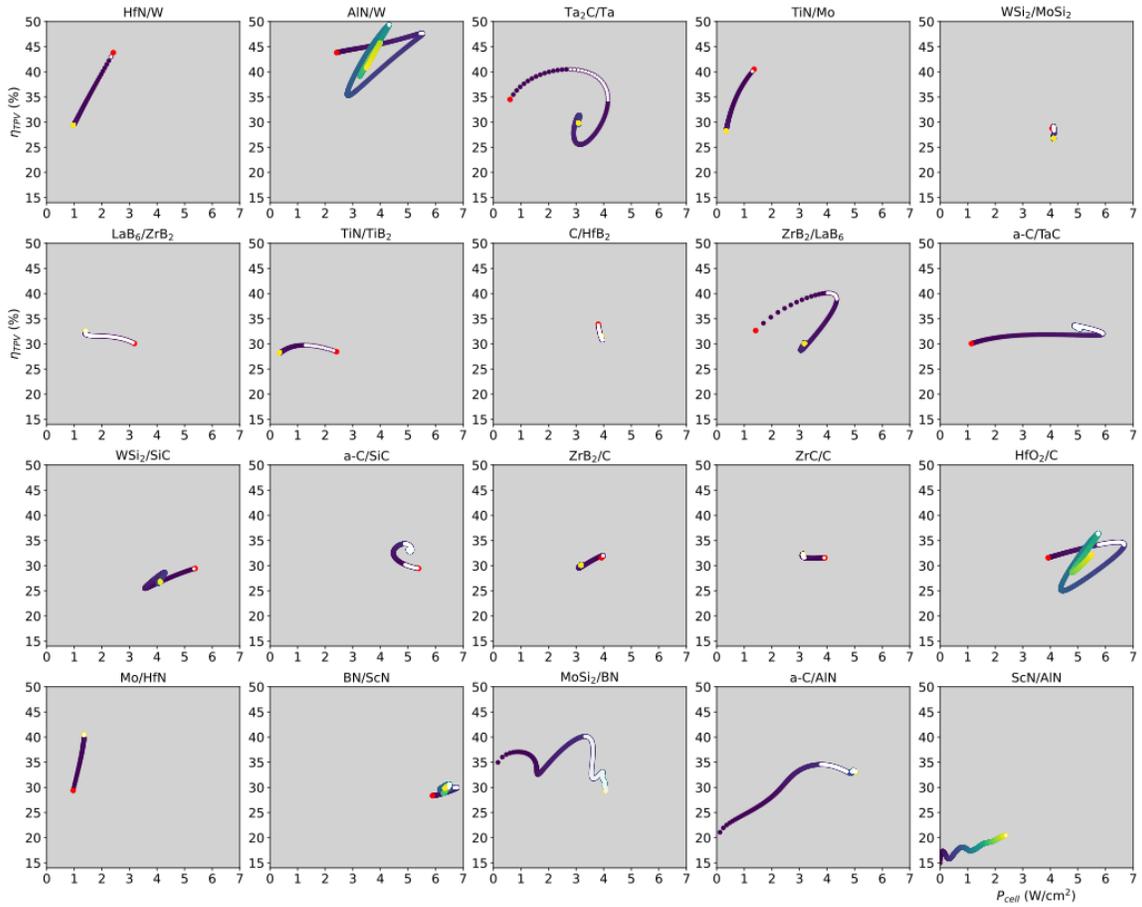

**Figure S7** Performance metric curves from Figure 2 all displayed with same performance metric space axes.

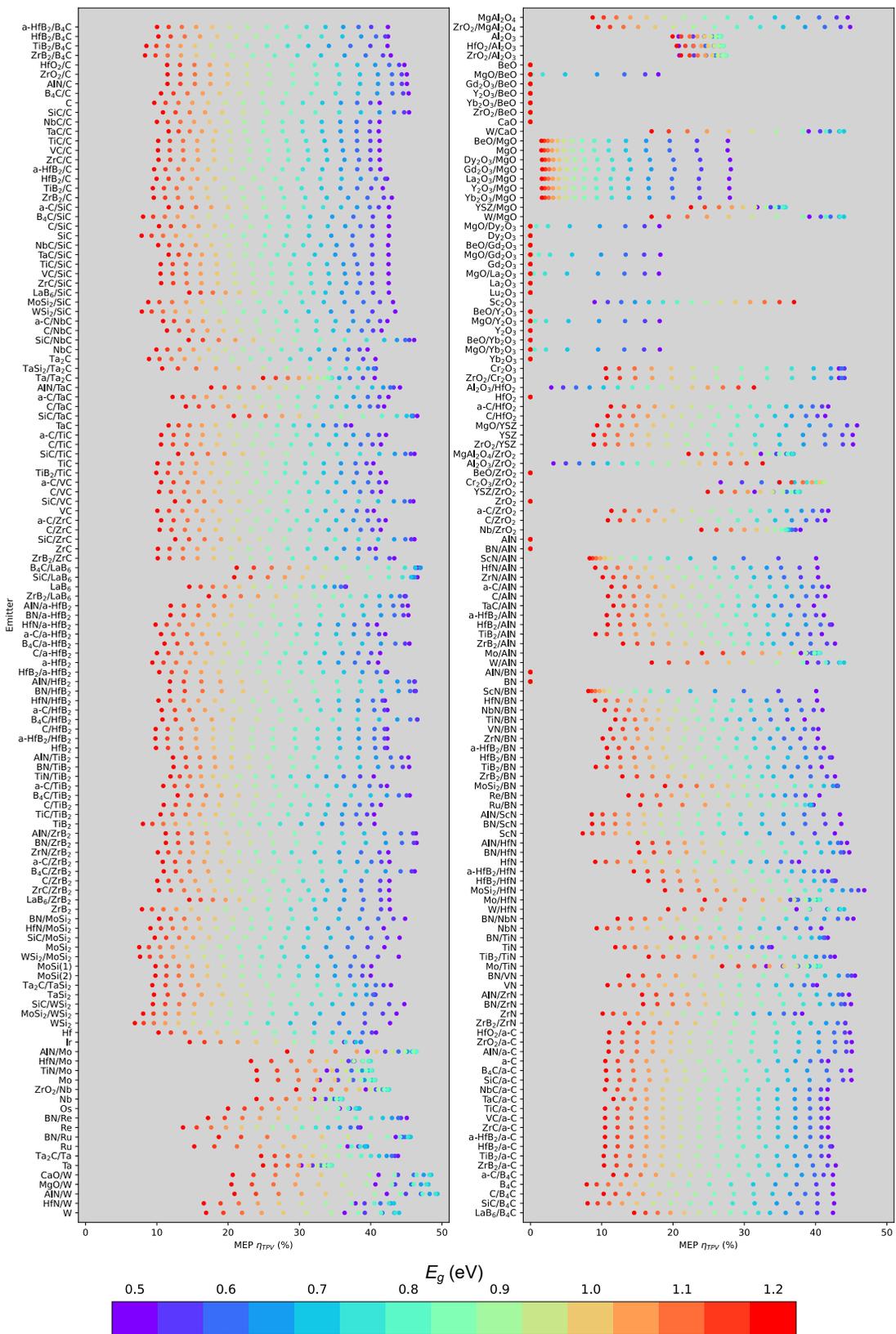

**Figure S8** MEP $\eta_{TPV}$ as a function of bandgap for 200 stable bilayer and 53 bulk emitters.

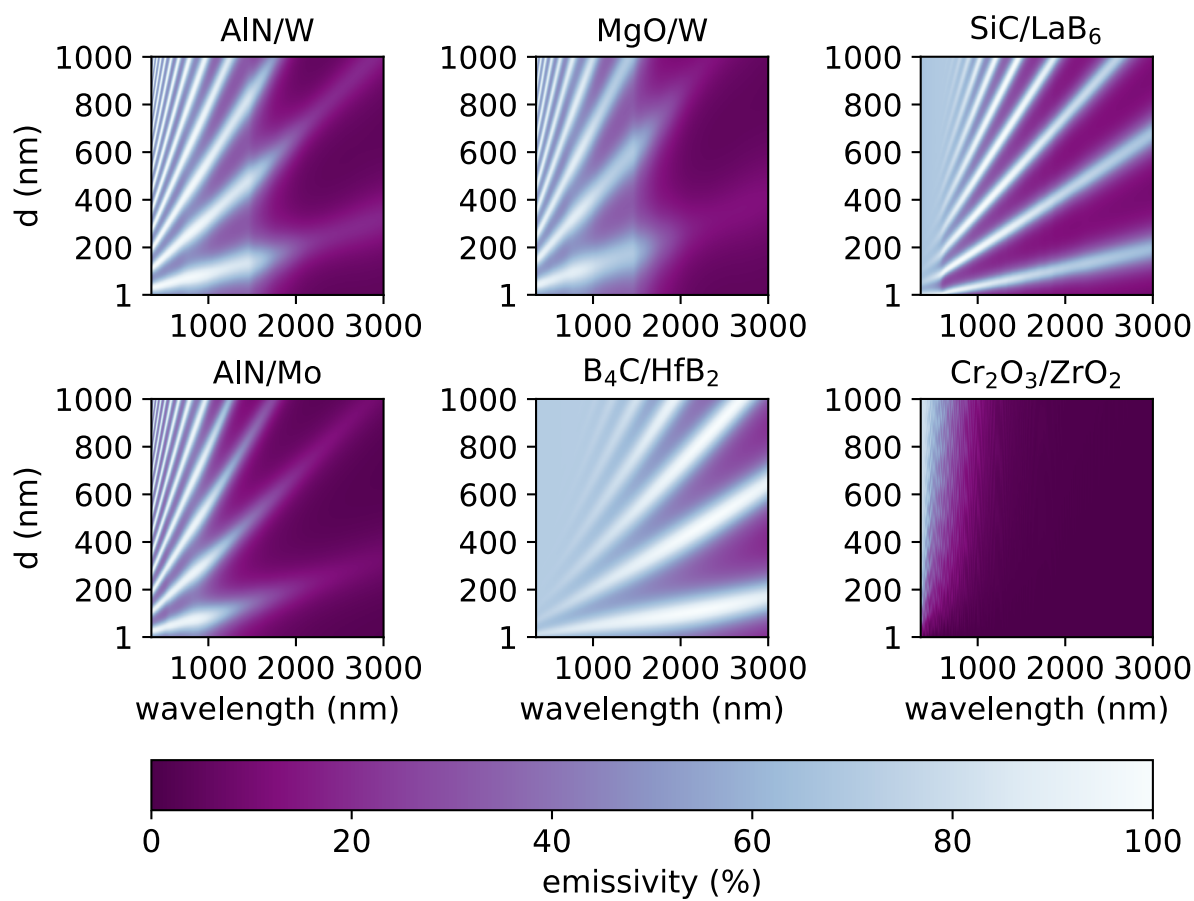

**Figure S9** Emissivity profiles from emitters Figure 3, where *d* refers to the thickness of the coating layer.